\documentclass[twocolumn]{svjour3}

\usepackage[utf8]{inputenc} 
\usepackage[T1]{fontenc}    
\usepackage{lmodern}
\usepackage{hyperref}       
\usepackage{url}            
\usepackage{booktabs}       
\usepackage{amsfonts}       
\usepackage{nicefrac}       
\usepackage{microtype}      
\usepackage{lipsum}
\usepackage{bm}
\usepackage{bbm}
\usepackage{upgreek}
\usepackage{mathtools}
\usepackage{natbib}
\usepackage{qtree}
\usepackage{color}
\usepackage{algorithm}
\usepackage{algorithmic}
\usepackage{amsmath}
\newcommand{\node}[2]{{$\left(\pi_{#1,#2}, \theta_{#1,#2}\right)$}}
\newcommand{\mockalph}[1]{}
\newcommand{\R}{{\mathbbm R}}
\newcommand{\I}{1{\hskip -2.5 pt}\hbox{I}}

\begin{document}

\title{Multiscale stick-breaking mixture models}

\author{Marco Stefanucci \and Antonio Canale}
\institute{Marco Stefanucci \at Department of Statistics, University of Padova \email{\texttt{stefanucci@stat.unipd.it}} \and Antonio Canale \at Department of Statistics, University of Padova \email{\texttt{canale@stat.unipd.it}}}

\maketitle

\begin{abstract}
	We introduce a family of multiscale stick-breaking mixture models for Bayesian nonparametric density estimation.  The Bayesian nonparametric  literature is dominated by single scale methods, exception made for P\`olya trees and allied approaches. Our proposal is based on a mixture specification exploiting an infinitely-deep binary tree of random weights that grows according to a multiscale generalization of a large class of stick-breaking processes; this multiscale stick-breaking is paired with specific stochastic processes generating sequences of parameters that induce stochastically ordered kernel functions. Properties of this family of multiscale stick-breaking mixtures are described. Focusing on a Gaussian specification, a Markov Chain Montecarlo algorithm for posterior computation is introduced. The performance of the method is illustrated analyzing both synthetic and real data sets. The method is well-suited for data living in $\mathbb{R}$ and is able to detect densities with varying degree of smoothness and local features.
\end{abstract}


\section{Introduction}

Nonparametric models have well-known advantages for their weak set of assumptions and great flexibility in a variety of situations. In particular, Bayesian nonparametrics (BNP) has received abundant attention in the last decades and it is nowadays a well-established modelling option in the data scientist's toolbox. If standard parametric Bayesian inference focuses on the posterior distribution obtained by defining suitable prior distributions over a finite dimensional parametric space $\Xi$ with $\xi \in \Xi$ typically characterising a specific parametric distribution $G_\xi$ for data $y = (y_1, \dots, y_n)$, in BNP one defines prior distributions on infinite-dimensional probability spaces flexibly characterizing the distribution $G$. Under these settings, only minor assumptions are made on $G$ making the whole inferential procedure more robust. 

The cornerstone of the discipline is the Dirichlet process (DP) introduced by \citet{Ferg73}. The DP is a stochastic process that defines a prior on the space of distribution functions; several generalizations of the DP have been proposed such as the Pitman--Yor (PY) process \citep{Per92, PY1997}, the normalized random measures with independent increments (NRMI)  \citep{igor2003, igor2004, igor2006, igor2009} and, more in general, the Gibbs-type priors  \citep{gnedin2006}. Realizations from these priors, however, are almost surely discrete probability functions and thus they do not admit a density with respect to the Lebesgue measure. As a remedy to this characteristic, the DP and allied priors can be used as prior distribution on the mixing measure of a mixture model. The first and most useful example of this is the DP mixture (DPM) of Gaussian kernels \citep{art:lo:1984, esco:west:1995}.  

\begin{figure*}[t]
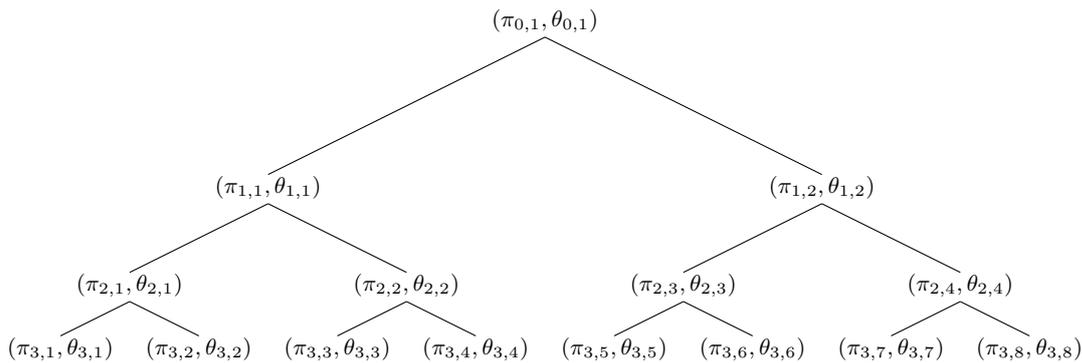

\Tree [.{$\left(\pi_{0,1}, \theta_{0,1}\right)$}
		[.{$\left(\pi_{1,1}, \theta_{1,1}\right)$}
			[.{$\left(\pi_{2,1}, \theta_{2,1}\right)$}
				[.\node{3}{1}  ] [.\node{3}{2}   ] ] 
			[.{$\left(\pi_{2,2}, \theta_{2,2}\right)$}
				[.\node{3}{3}  ] [.\node{3}{4}   ] ] ]
		[.{$\left(\pi_{1,2}, \theta_{1,2}\right)$}
			[.{$\left(\pi_{2,3}, \theta_{2,3}\right)$}
				[.\node{3}{5}  ] [.\node{3}{6}   ] ] 
			[.{$\left(\pi_{2,4}, \theta_{2,4}\right)$}
			[.\node{3}{7}  ] [.\node{3}{8}   ] ] ]
	]
\caption{Binary tree with mixture weights $\pi_{s,h}$ and kernel's parameters $\theta_{s,h}$ at each node $(s, h)$, where $s$ is the scale level and $h$ is the index within the scale.}\label{fig:tree}
\end{figure*}
\setcounter{figure}{1}

P\'olya trees (PT) \citep{lavi:1992a,lavi:1992b,maul:etal:1992} are alternative formulations whose draws implicitly admit densities with respect to the Lebesgue measure. PT, on the surface, are also particularly appealing in providing a multiscale structure and thus in characterizing possible abrupt local changes on the density. In practice, however, this construction tends to produce highly spiky density estimates even when the true density is smooth. A recent alternative formulation based on PT that circumvents this lack of smoothness is the smoothed PT prior proposed by \citet{cipolli}. This prior distribution can naturally model continuous densities on $\R$. \citet{canale:sinica} consider a related multiscale mixture model based on Bernstein polynomials but are confined to model continuous densities on $(0,1)$.

Consistently with these contributions, in this paper we introduce a general class of multiscale stick-breaking processes with support on the space of discrete probability mass functions suitable as mixing measure in multiscale mixture of continuous kernels. The method generalizes \citet{canale:sinica} in two directions. First, a more general multiscale stick-breaking process inspired by the PY process is introduced. Second, a multiscale base measure generating kernel densities that are stochastically ordered and defined on a general sample space is introduced. This construction leads to a class of prior measure for continuous densities that is robust to any specific prior parameters elicitation and that naturally adapts to the actual degree of smoothness of the true data generating distribution without the need of specifying several layers of hyper-priors. 

The remainder of the paper is organized as follows. In the next section we introduce our multiscale stick-breaking prior and  describe some of its properties. Section \ref{sec:computation} describes a Gibbs sampling algorithm for posterior computation. Section \ref{sec:illustration} illustrates the performance of the methods through the analysis of several synthetic and real datasets. Section \ref{sec:end} concludes the paper.

\section{Multiscale stick-breaking mixture}

Let $y \in \mathcal{Y} \subset \R$, be a random variable with unknown density $f$. 
We assume for $f$ the following multiscale construction 
\begin{equation}
\label{eq:1}
    f(y)= \sum_{s=0}^\infty \sum_{h=1}^{2^s} \pi_{s,h} \mathcal{K}(y;{\theta}_{s,h}),
\end{equation}
where $\mathcal K(\cdot; \theta)$ is a kernel function parametrized by $\theta \in \Theta$ and $\{ \pi_{s,h}\}$ and $\{{\theta}_{s,h}\}$ are unknown sequences of positive weights summing to one and  parameters belonging to $\Theta$, respectively. 
We will refer to this model with the term multiscale mixture (MSM) of kernel densities.
This construction can be represented with an infinitely deep binary tree in which each node is indexed by a scale $s$ and an index $h = 1, \dots, 2^s$ and where each of these nodes is characterized by the pair $(\pi_{s,h}, \theta_{s,h})$. A cartoon of a truncation of this binary tree is reported in Figure~\ref{fig:tree}.

Model \eqref{eq:1} can be equivalently written as
\begin{equation}
\label{eq:2}
f(y)= \int \mathcal{K}(y;{\theta}) dP(\theta), \quad P = \sum_{s=0}^\infty \sum_{h=1}^{2^s} \pi_{s,h} \updelta_{\theta_{s,h}},
\end{equation}
where $\updelta_x$ is the Dirac delta function. Thus a prior distribution for the multiscale mixture \eqref{eq:1} is obtained by specifying suitable stochastic processes for the random mixing measure $P$ or, equivalently, for the random sequences $\{\pi_{s,h}\}$ and $\{\theta_{s,h}\}$. These characterizations are separately carefully described in the next sections.

Approximations of the mixture model \eqref{eq:1}  can be obtained fixing an upper bound $s'$ for the depth of the tree. This truncation is obtained setting $S_{s'} =  1$ for each $h = 1,\dots ,2^{s'}$ as suggested by \citet{lancillotto} for the standard single-scale mixture model and discussed by \citet{canale:sinica} for the multiscale mixture of Bernstein polynomial model. Such a truncation can be applied both if one considers not scientifically relevant higher levels of resolution or to reduce the computational burden.

\subsection{Multiscale mixture weights}
\label{sec:stickbreak}

We first focus on the sequence of mixture weights $\{\pi_{s,h}\}$. We introduce independent random variables $S_{s,h}$ and $R_{s,h}$ taking values in $(0,1)$ and describing the probability of taking a given path in the binary tree reported in Figure~\ref{fig:tree}. Specifically, $S_{s,h}$ denotes the probability of stopping at node $h$ of scale $s$ while $R_{s,h}$ denotes the probability of taking the right path from scale $s$ to scale $s+1$ conditionally on not stopping in node $h$ of that scale. The weights are then defined as
\begin{equation}
\pi_{s,h} = S_{s,h} \prod_{r<s} (1-S_{r,\lceil h2^{r-s} \rceil}) T_{shr},
\label{eq:weights}
\end{equation}
where  $T_{r, \lceil h2^{r-s} \rceil}  = R_{r, \lceil h2^{r-s} \rceil}$, 
if $(r+1,\lceil h2^{r-s+1} \rceil)$ is the right daughter of node $(r,\lceil h2^{r-s} \rceil$, and 
  $T_{r, \lceil h2^{r-s} \rceil} = 1-R_{r, \lceil h2^{r-s} \rceil}$, otherwise.
This construction is reminiscent of the stick-breaking process \citep{art:seth:1994, lancillotto} and can be described by the following metaphor:
 take a stick of length one and break it according to the law of $S_{0,1}$; the remainder of the stick is then randomly splitted in two parts according to the law of $R_{0,1}$; at general node $(s,h)$ the remainder of the stick, conditionally on the previous breaks, is broken according to $S_{s,h}$ and then splitted according to  $R_{s,h}$. 
 
 Different distributions for $S_{s,h}$ and $R_{s,h}$ lead to different characteristics for the tree of weights. Inspired by the general stick-breaking prior construction of \citet{lancillotto} we can set
\begin{equation}
S_{s,h} \sim \mbox{Be}(a_{s,h}, b_{s,h}),
\quad  
R_{s,h} \sim \mbox{Be}(c_{s,h}, d_{s,h}).
\label{eq:generalstickbreaking}
\end{equation}

This  construction is a  flexible  generalization of  \citet{canale:sinica} that, mimicking the DP and its stick-breaking representation,  fixed $a_{s,h} = 1$, $b_{s,h} = \alpha > 0 $, and $c_{s,h} = d_{s,h} = \beta > 0 $ for each $s$ and $h$. While being way more flexible, the specification in \eqref{eq:generalstickbreaking} has different parameters for each node and its elicitation may be cumbersome in practice. To avoid these complications while  keeping  an increasing degree of flexibility through a scale dependence for the distribution of the random weights, we consider  $\delta \in [0,1)$ and $\alpha> -\delta$ and let
\begin{equation}
S_{s,h} \sim \mbox{Be}(1-\delta, \alpha + \delta (s+1) ),
\quad  
R_{s,h} \sim \mbox{Be}(\beta, \beta).
\label{eq:SR}
\end{equation}
This specification is  reminiscent of the PY process,  a model which stands out for being a good compromise between modelling flexibility, and mathematical and computational tractability. This  construction leads to a proper sequence of weights as formalized in the next Lemma. Its proof is reported in the Appendix.

\begin{lemma}
	\label{lem:sumtoone}
	Let $\pi_{s,h}$ be an infinite sequence of weights defined by  \eqref{eq:weights} and \eqref{eq:SR}. Then, for any $\beta>0$, $\delta \in [0,1)$, and $\alpha >-\delta$
	\begin{equation}
	\sum_{s=0}^\infty \sum_{h=1}^{2^s} \pi_{s,h} = 1 
	\label{eq:sumtoone}
	\end{equation}
	almost surely.
\end{lemma}

 The $\delta$ parameter allows for a greater degree of flexibility in describing how the random probability weights are allocated to the nodes. To see this, consider the expectation of $\pi_{s,h}$, i.e.
 \begin{eqnarray*}
 \mathbb{E}(\pi_{s,h}) 
 & = &  \mathbb{E}\bigg\{ S_s \prod_{l=0}^{s-1} (1-S_l) \prod_{l=1}^s T_l \bigg\} \\
 & = & 
 \bigg( \frac{ 1-\delta  }{ \alpha +1 } \bigg) \bigg( \frac{1}{2} \bigg)^s \prod_{l=1}^s \bigg(\frac{ \alpha +\delta l }{  \alpha +\delta l + 1 } \bigg),
\label{eq:mean}
 \end{eqnarray*}
  where we discard the $h$ subscript on $S_l \sim \mbox{Be}(1-\delta , \alpha + \delta (l +1))$ and $T_l \sim \mbox{Be}(\beta,\beta)$ for ease in notation.  This does not impact the calculation because any path taken up to scale $s$ has the same probability {\em a priori} and the distribution of the random variables in \eqref{eq:SR} depends on the scale $s$ only. The expected values of the random weights can be used to calculate the expected scale at which an observation falls, a measure of the expected resolution level, defined by $\mathbb{E}(\tilde{S}) = \sum_{s=0}^\infty s  \mathbb{E}(\pi_{s,h}) $. The latter  simplifies to $\alpha$ when $\delta = 0$  but can be easily obtained numerically for $\delta> 0$.

\begin{figure*}[t] 
	\centering
	\includegraphics[width=0.9\textwidth]{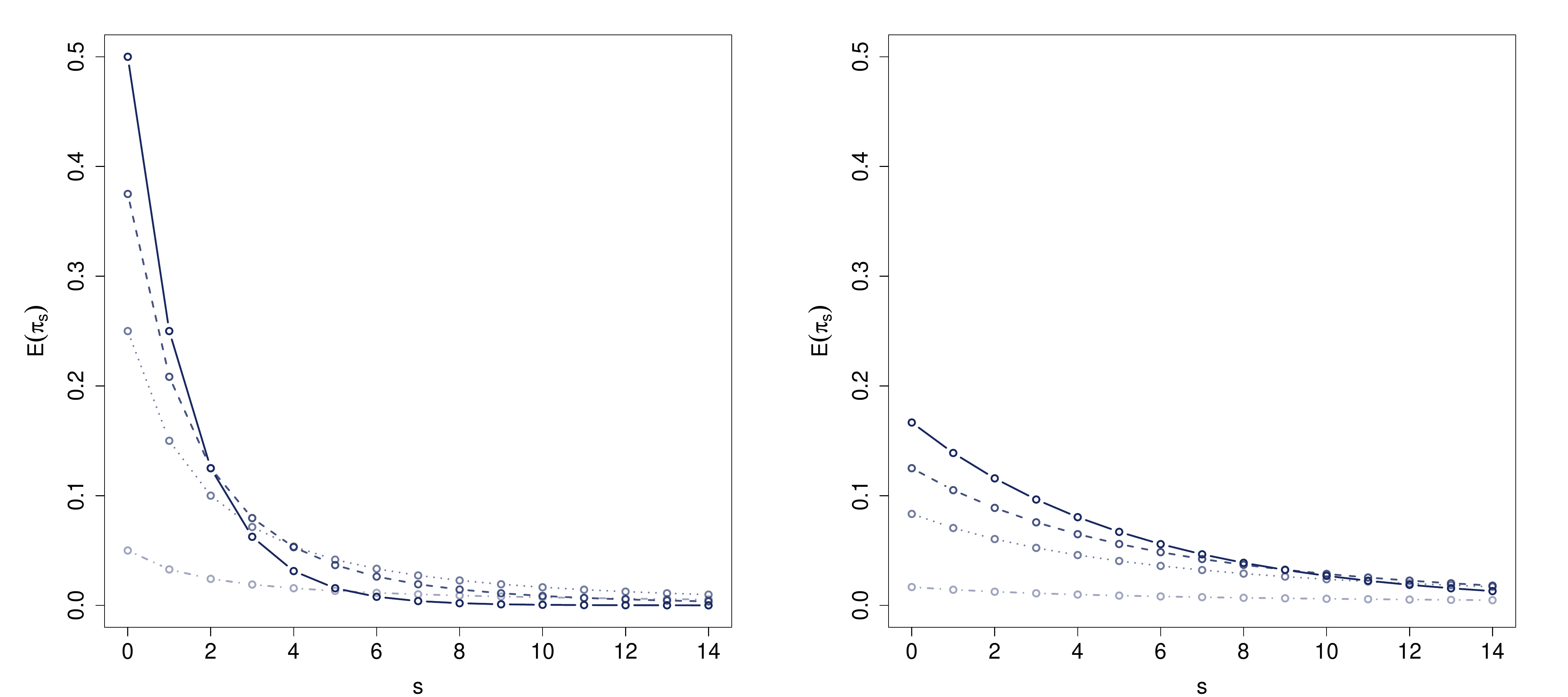} 
	\caption{Prior total weight $\pi_s = \sum_{h<2^s} \pi_{s,h}$ as a function of $s$ and for $\delta$ equal to $0$ (---), 0.25 ($- -$), 0.5 ($\cdots\cdot$), and 0.9 ($\cdot - \cdot -$); $\alpha=1$ (left) and $\alpha = 5$ (right).
	}
	\label{fig:pesidelta}
\end{figure*}

 To better understand the role of $\delta$, Figure~\ref{fig:pesidelta} reports the  total expected  weight of scale $s$, defined as the expectation of $\pi_s = \sum_{h} \pi_{s,h}$, for different values of $\delta$ and $\alpha$. It is clear that increasing values of $\delta$ make the first levels of the tree less probable \emph{a priori}, thus favoring a deeper tree. Note that this characteristic has to be interpreted more in terms of prior robustness rather than favouring rougher densities as the prior mass is more spread through the whole tree allowing the posterior to concentrate on a tree of suitable depth. This interpretation is consistent with the role of the discount parameter of the PY process that controls  how much prior probability is concentrated around the prior expected value of occupied clusters and thus inducing a posterior distribution that is more robust to the prior specification. See \citet{ieee2015} for a related discussion.  We will show in  Section \ref{sec:illustration1} that this conjecture is empirically confirmed  via simulations.

\subsection{Multiscale kernel's parameters}

We now discuss the stochastic process  for the sequence $\{\theta_{s,h}\}$. For ${\cal Y} = (0,1)$, \citet{canale:sinica} assume that $\mathcal K(\cdot;\theta_{s,h})$ is a Beta($h, 2^s-h+1$) density so that $\theta_{s,h}$ is identified by the pair $(h, 2^s-h+1)$, a fixed set of parameters. 
This construction is implicitly inducing a mixture of Bernstein polynomials \citep{petr:1999a,petr:1999b} for each scale $s$ and the randomness is totally driven by the sequence of mixture weights. Here, instead, we will consider the broader case where  $\theta_{s,h}$ are unknown parameters and where ${\cal K}(\cdot; \theta)$ is a location-scale kernel defined on a general sample space $\mathcal Y$. Under this specification, we partition the kernel's parameter space into a location and scale part letting $\Theta = \Theta_{\mu} \times \Theta_{\omega}$. 

\subsubsection{Location parameters}
\label{sec:locations}

We first focus on defining a suitable sequence of locations $\{\mu_{s,h}\}$ that, consistently with the dyadic partition induced by the binary tree structure, uniformly covers the  space $\Theta_{\mu}$. To this end, for any scale $s$ we introduce a partition of $\Theta_{\mu}$ by letting
\begin{equation}
\label{eq:partition1}
\Theta_{\mu}  = \bigcup_{h=1}^{2^s} \Theta_{\mu;s,h} ,
\end{equation}
such that for two neighboring scales $s$ and $s+1$,
\begin{equation}
\label{eq:partition2}
\Theta_{\mu;s,h} = \Theta_{\mu;s+1,2h-1} \cup \Theta_{\mu;s+1,2h}.
\end{equation}
Let  $G_{0}$ be a base probability measure defined on $\Theta_{\mu}$ and use it both to define $\Theta_{\mu;s,h}$ and to generate the multiscale locations  $\mu_{s,h}$. Specifically, we  set
\begin{equation}
\label{eq:partitionquantiles}
\Theta_{\mu;s,h} = \big[q_{\frac{h-1}{2^s}}, q_{\frac{h}{2^s}}\big],  
\end{equation}
where $q_{r}$ is the $r$-level quantile of the density of $G_0$. Then random $\mu_{s,h}$ are sampled proportionally to $G_{0}$ truncated in  $\Theta_{\mu;s,h}$.  While preserving the covering of the $\Theta_{\mu}$ this construction allows for straightforward prior elicitation similarly to what is done for the DP or the PY process. The next lemma, whose proof is reported in the  Appendix, shows that \emph{a priori} the random probability measure on $\Theta_\mu$ defined  by
\begin{equation}
G =   \sum_{s=0}^\infty \sum_{h=1}^{2^s} \pi_{s,h} \updelta_{\mu_{s,h}}
\label{eq:Gmu}
\end{equation}
is centered around $G_0.$

\begin{lemma}
	\label{lem:cntrd}
	Let $G_{0}$ be a base probability measure defined on $\Theta_{\mu}$. Introduce a dyadic recursive partition of $\Theta_{\mu}$ defined  by \eqref{eq:partition1}, \eqref{eq:partition2}, and \eqref{eq:partitionquantiles} and $G_0$. If $G$ is the discrete measure \eqref{eq:Gmu} and each  $\mu_{s,h}$ is randomly sampled proportionally to $G_0$ truncated in $\Theta_{\mu;s,h}$, then, for any set $A \subseteq \Theta_{\mu}$,
	\begin{equation*}
	\mathbb{E}[G(A)] = G_{0}(A).
	\end{equation*}
\end{lemma}

Note that equation \eqref{eq:Gmu} is similar, in spirit,  to the approximate PT (APT) prior of \citet{cipolli} and in particular to their equation (3). The difference, however, is twofold. First, our weights come from a multiscale stick-breaking process while those of APT are the result of the PT recursive partitioning. The second, more evident, difference lies on how the Dirac's delta masses are placed. While equation (3) of \citet{cipolli} places these on the center of the intervals $\Theta_{\mu; s,h}$, in our construction the masses are randomly placed inside $\Theta_{\mu; s,h}$. Hence, while the learning in APT model is totally driven by the random weights,  our approach allows for an update of the values $\mu_{s,h}$ \emph{a posteriori}.

\subsubsection{Scale  parameters}
\label{sec:scales}

We now focus on describing the sequence of scale parameters $\{\omega_{s,h}\}.$ Consistently with our multiscale setup, the scale parameters need to be ordered with respect to the scale levels of the binary tree in order to induce more concentrated kernels for increasing values of $s$, on average. In general the direction of the ordering depends on the actual role of the scale parameters in the specific kernel ${\cal K}(\cdot; \theta)$. For instance, for scale parameters proportional to the variances---respectively precisions--- a decreasing---respectively increasing---sequence need to be specified. Assuming that $\omega_{s,h}$ are proportional to the variances of the kernels,  we induce a stochastic ordering of the $\omega_{s,h}$'s at different scales $s$ in the following way. 
Let $H_0$ be a base probability measure defined on $\Theta_{\omega}$ with first moment $\mathbb{E}_{H_0}(\omega) = \omega_0$ and variance  $\mathbb{V}_{H_0}(\omega) = \gamma_0$ both finite. Then let
\begin{equation}
\omega_{s,h} =  c(s) W_{s,h}, \quad W_{s,h} \stackrel{iid}{\sim} H_0,
\label{eq:scales}
\end{equation}
where $c(s)$ is a monotone decreasing deterministic function of $s$. Under this definition 
the sequence of $\{\omega_{s,h}\}$ is stochastically decreasing and
\[
\mathbb{E}_{H_0}(\omega_{s+1,h}) \leq \mathbb{E}_{H_0}(\omega_{s,h}) , \quad 
\mathbb{V}_{H_0}(\omega_{s+1,h}) \leq \mathbb{V}_{H_0}(\omega_{s,h}).
\]
Consistently with our multiscale construction, the first inequality reflects the fact that from scale $s$ to scale $s+1$ we expect more concentrated kernels in equation \eqref{eq:2}. The second inequality, in addition, implies that the prior uncertainty about $\omega$ scales as well. 

In the next section we discuss a specification of this construction by means of Gaussian kernels and suitable choices for $G_0$, $H_0$, and $c(s)$.

\subsection{Multiscale mixture of Gaussians}
\label{sec:mixgaussian}
Although several choices for the kernel $\mathcal K(\cdot; \theta)$ can be made, the Gaussian one is probably the more natural when ${\cal Y} = \R$. Hence we specify the model described in previous sections assuming 
${\cal K}(\cdot; \theta) = \phi(\cdot; \mu, \omega)$ where $\phi(\cdot; \mu, \omega)$ is a Gaussian density with mean $\mu\in \R$ and variance $\omega>0$. Under this specification equation \eqref{eq:1} becomes a MSM of Gaussian densities, i.e.
\begin{equation*}
    f(y)= \sum_{s=0}^\infty \sum_{h=1}^{2^s} \pi_{s,h} \phi(y;\mu_{s,h},\omega_{s,h}).
\end{equation*}

A pragmatic choice for the base measures consists in choosing conjugate priors. Specifically we let $G_0$ be a Gaussian distribution with mean $\mu_0$ and variance $\kappa_0$. Similarly, we restrict to the inverse-gamma family of distributions the choice for $H_0$.  Following \eqref{eq:scales}, we let 
$
W_{s,h} \stackrel{iid}{\sim} \mbox{IGa}(k, \lambda),
$
leading to  $\mathbb{E}(W_{s,h})= \lambda/(k-1)$ and  $\mathbb{E}(\omega_{s,h})= c(s) \lambda/(k-1)$.  A natural choice for the function $c(\cdot)$ is $c(s)=2^{-s}$, which is equivalent to let
$
\omega_{s,h} \sim \mbox{IGa}(k, 2^{-s}\lambda).
$

Consistently with the discussion at the end of Section  \ref{sec:locations}, this final specification is reminiscent of the smoothed approximate Polya tree (SAPT) of \citet{cipolli}. In both specifications, indeed,  the variances of each Gaussian mixture component are the result of a deterministic scale-decreasing component---represented by the function $c(s)$ here and by the parameter $d_k$ in SAPT---and a random quantity. The latter, while being controlled by a single parameter in the SAPT model, is component-specific in the proposed formulation thus allowing for local learning of the values of each scale parameter.  

\section{Posterior Computation}
\label{sec:computation}

In this section we introduce a Markov Chain Montecarlo (MCMC) algorithm to perform posterior inference under the model introduced in the previous section. In the general settings, the algorithm consists of three steps: (i) allocate each observation 
to a multiscale cluster conditionally on the current values of  $\{ \pi_{s,h} \}$ and   $\{ \theta_{s,h} \}$; 
(ii) update $\{ \pi_{s,h} \}$ conditionally on the cluster allocations; 
(iii) update $\{ \theta_{s,h} \}$ conditionally on the cluster allocations.

In this section we focus on the multiscale mixture of Gaussian and related prior elicitation discussed in Section \ref{sec:mixgaussian} but steps (i) and (ii) also apply for a general kernel.

Suppose subject $i$ is assigned to node $(s_i,h_i)$, with $s_i$ the scale and $h_i$ the node within scale.
Conditionally on the values of the parameters, the posterior probability of subject $i$ belonging to node ($s,h$) is simply
\begin{align*}
	\mathbb{P}(s_i = s, h_i=h | y_i, \pi_{s,h}) & \propto \pi_{s,h} {\cal K}(y; \theta_{s,h}).
\end{align*}
Consider the total mass assigned at scale $s$, defined as $\pi_s = \sum_{h=1}^{2^s} \pi_{s,h}$, and let $\bar{\pi}_{s,h} = \pi_{s,h}/\pi_{s}$. Under this notation, we can rewrite the likelihood for the $i$-th observation as 
\begin{align*}
	f(y_i) = \sum_{s=0}^\infty \pi_{s} \sum_{h=1}^{2^s} \bar{\pi}_{s,h}  {\cal K}(y_i; \theta_{s,h}).
\end{align*}
Following \citet{slice},  we  introduce the auxiliary random variables 
$
	u_i | y_i, s_i \sim \mbox{Unif}(0, \pi_{s_i}), 
 $
and consider the joint density 
\[
f(y_i,u_i,s_i) \propto \I_{(0, \pi_{s_i})}(u_i) \sum_{h=1}^{2^{s_i}} \bar{\pi}_{s_i,h}  {\cal K}(y_i; \theta_{s,h}),
\]
where $\I_A(x)$ is the indicator function that returns 1 if $x \in A$.
Then we can update the scale $s_i$ and the node $h_i$ using 
\begin{align*}
	& \mathbb{P}(s_i=s \mid u_i,y_i) \propto \I_{[u_i,1]} (\pi_s)\sum_{h=1}^{2^{s}} \bar{\pi}_{s,h}  {\cal K}(y_i; \theta_{s,h}), \\
	& \mathbb{P}(h_i=h | u_i,y_i,s_i) \propto \bar{\pi}_{s_i,h} {\cal K}(y_i; \theta_{s,h}).
\end{align*}


%

Conditionally on cluster allocations, the update of the weights is obtained applying \eqref{eq:weights} to the updated values of $S_{s,h}$ and $R_{s,h}$ obtained sampling from 

\begin{align*}
S_{s,h} &\sim \mbox{Be}(1-\delta + n_{s,h}, \alpha + \delta (s+1)  + v_{s,h} - n_{s,h}), \\
	R_{s,h} & \sim \mbox{Be}(\beta+r_{s,h}, \beta + v_{s,h}  - n_{s,h} - r_{s,h} ), 
\end{align*}
where $v_{s,h}$ is the number of subjects passing through node $(s,h)$, $n_{s,h}$ is the number of subjects stopping at node $(s,h)$,   and  $r_{s,h}$ is the number of subjects that continue to the right after passing through node $(s, h)$.

Conditionally on cluster allocation, the update of locations and scale parameters follows from usual conjugate analysis arguments. Specifically the location parameters are sampled from 
\begin{equation*}
	\mu_{s,h} \sim  \text{N}_{\Theta_{\mu;s,h}} 	\biggl(
\frac{ \mu_0\omega_{s,h}+ n_{s,h} \bar{y}_{s,h}\kappa_0}{ n_{s,h}\kappa_0 + \omega_{s,h}},
\frac{\omega_{s,h}\kappa_0}{ n_{s,h}\kappa_0 + \omega_{s,h}} \biggl), 
\end{equation*}
where $\bar{y}_{s,h}$ is the sample mean of the observations assigned to node $(s,h)$, and $N_A(m,v)$ denotes a Gaussian distribution with mean parameter $m$ and variance parameter $v$ truncated in the set $A$. The scale parameters are sampled from 
\begin{equation*}
	\omega_{s,h} \sim \mbox{IGa}\biggl(k +\frac{n_{s,h}}{2}, \frac{\lambda}{2^s} + \frac{\sum_{i: s_i = s, h_i = h}(y_i-\mu_{s,h})^2}{2}\biggl).
\end{equation*}

\section{Illustrations}
 \label{sec:illustration}
 
 In this section we discuss the performance of the proposed MSM of Gaussian densities through the analysis of different synthetic and real data sets.  Specifically, we investigate the role of the  $\delta$  parameter in the next section and compare the method with the DPM and the SAPT in Section \ref{sec:comparrison}. Finally, in Sections \ref{sec:galaxy1} and \ref{sec:galaxy2} the method and one possible extension of it are used to analyze two different astronomical data sets.
 
\subsection{The role of $\delta$}
 \label{sec:illustration1}
 
As already discussed in the previous sections, the $\delta$ parameter allows for a greater degree of flexibility in the prior specification. In this section we want to empirically assess its role \emph{a posteriori}. To this end  we generate 100  samples of size $n=50$  from  three different densities and run the Gibbs sampling algorithm described in Section~\ref{sec:computation} to get an estimate of the posterior mean density for different values of the $\alpha$ and $\delta$ parameters. 

Data are generated from a finite mixture of  Gaussian densities $f(y) = \sum_{k=1}^K \pi_{k} \phi(y;\mu_k,\omega_k)$ with an increasing level of local variability. Specifically, the first density is the standard normal distribution, the second density is a mixture  of two components with $\mu_1=-\mu_2 = 0.935$, $\omega_1=\omega_2=1/8$, and $\pi_1=\pi_2=1/2$, while the last density has three components and parameters equal to $\mu_1=0$, $\mu_2=1.392$, $\mu_3=-1.392$, $\omega_1=\omega_2=\omega_3=1/32$, $\pi_1=1/2,\pi_2=\pi_3=1/3$. 

We considered $\delta$ equal to $0, 0.25$ and $0.5$ and numerically obtain the values of  $\alpha$ in order to match a fixed prior expectation for the scale of the density. We considered three values for the prior expected scale that are consistent with the densities of the data generating processes. Specifically we assume $\mathbb{E}(\tilde{S})  = 1, 3$ and 5. The related parameters are summarized in Table~\ref{tab:priorsims}.
\begin{table}[h]
		\caption{Values of  $\alpha$ parameters for given $\delta$ and expected scales $\mathbb{E}(\tilde{S})$.}
		\centering
	\begin{tabular}{cc|rrr}
		 && \multicolumn{3}{c}{$\mathbb{E}(\tilde{S})$}\\
		& & \multicolumn{1}{c}{1}& \multicolumn{1}{c}{3}& \multicolumn{1}{c}{5}\\
		\hline
		& 0.00& 1.00& 3.00& 5.00\\
	$\delta $ &  0.25& 0.25& 1.25& 2.25\\		
	& 0.50& -0.45& -0.35& -0.25\\
\end{tabular}
\label{tab:priorsims}
\end{table}

We run the Gibbs sampler described in Section \ref{sec:computation} for 1000 iterations with a burn in of 200. Visual inspections of the traceplots of the posterior mean density on a grid of domain points suggests no lack of convergence. 

\begin{figure*}[t] 
	\centering
	\vspace*{0cm}
	\includegraphics[width=1\textwidth]{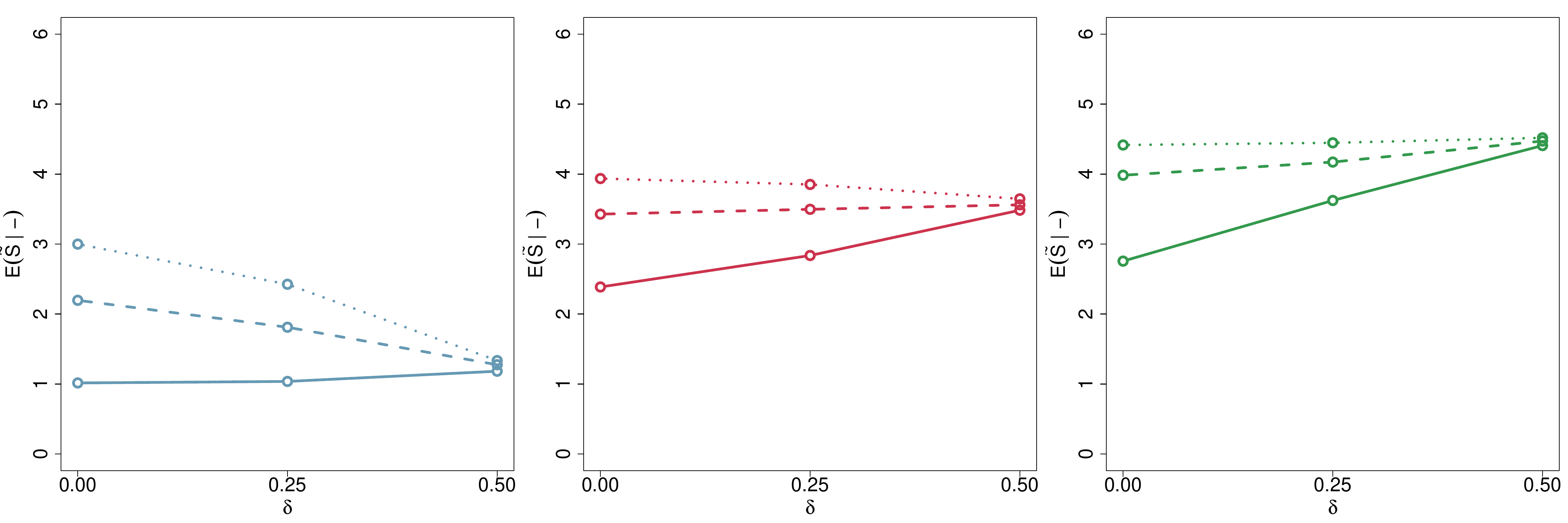} 
	\caption{Posterior scale as a function of $\delta$ for different values of $\mathbb{E}({\tilde S})$. Continuous line: $\mathbb{E}({\tilde S})=1$; dashed line: $\mathbb{E}({\tilde S})=3$; dotted line: $\mathbb{E}({\tilde S})=5$; First plot: Standard normal distribution. Second plot: mixture of two Gaussians (see text). Third plot: mixture of three Gaussians (see text). Sample size equal to 50.
	}
	\label{fig:internalsim}
\end{figure*}	

Figure~\ref{fig:internalsim} reports the values of the average (over the 100 replicates) of the posterior mean scale as a function of the discount parameter $\delta$. Each dot corresponds to a specific configuration of the $\alpha$ and $\delta$ parameters and configurations with the same prior mean scale are connected. For $\delta =0$  the prior choice drives the behavior of the posterior, i.e. on average lower posterior mean is obtained when the prior mean scale is equal to 1 while a higher posterior mean is obtained when the prior mean is equal to 5. This prior dependence is less evident for increasing values of $\delta$. Indeed regardless the prior specification, when the value of $\delta$ increases, the posterior mean stabilizes in a neighborhood of a specific value. This behavior is consistent with what happens for the posterior mean number of clusters for a PY process mixure model \citep[see][for related discussions]{ieee2015, cana:prue}. 

Note that in addition to this \emph{prior robustness} on the posterior mean scale---that is related to the actual degree of smoothness of the posterior mean density---we also observe an increasing precision of the density estimates in terms of $L_1$ distance  of the posterior mean density and the true density. See the Supplementary Materials for additional details. 

The same simulation experiment was carried out also for data sets with sample size equal to $250$. The qualitative results are similar but less striking as the different posterior mean scales are closer for small values of $\delta$. This is expected and reflects the informative gain related to a bigger sample size. Additional details and plots are reported in the Supplementary Materials.

 \subsection{Comparison with alternative methods}
\label{sec:comparrison}

\begin{table*}[t]
	\centering
	\caption{ Mean and standard deviation ($\times10^3$) of the $L_1$ distance and KL divergence between the estimated posterior density and the true data generating density over 100 simulations.}
	\begin{tabular}{crrrrrr}\toprule
		& MSM  &  & DPM &  & SAPT & \\
			\cline{2-7}\\
		& \multicolumn{1}{c}{$L_1$} & 
		\multicolumn{1}{c}{\textit{KL}} & 
		\multicolumn{1}{c}{$L_1$} & 
		\multicolumn{1}{c}{\textit{KL}} & 
		\multicolumn{1}{c}{$L_1$} &
		\multicolumn{1}{c}{\textit{KL}} \\
	\cline{2-3}
		\cline{4-5}
		\cline{6-7}\\
		{\small $n=100$}\\
		\vspace*{.1cm}
		S1 &  150.82 \scriptsize {(47.42)}& 26.13 \scriptsize{(13.98)} & 163.57 \scriptsize{(55.91)}& 32.90 \scriptsize{(15.29)} & 214.46 \scriptsize{(52.61)} & 58.87 \scriptsize{(31.19)}  \\
		\vspace*{.1cm}
		S2   & 178.90  \scriptsize{(48.54)}& 32.51 \scriptsize{(14.95)}& 214.34 \scriptsize{(60.47)}&  47.04 \scriptsize{(22.14)} & 203.40 \scriptsize{(46.88)} & 41.69 \scriptsize{(17.01)}   \\
		\vspace*{.1cm}
		S3   & 288.27  \scriptsize{(38.24)}& 89.35 \scriptsize{(12.86)}& 319.60 \scriptsize{(34.41)}&  105.65 \scriptsize{(12.99)} & 285.97 \scriptsize{(43.78)} & 93.79 \scriptsize{(15.83)}  \\
		\vspace*{.1cm}
		S4   & 230.52  \scriptsize{(42.87)}& 47.77 \scriptsize{(14.29)}& 243.98 \scriptsize{(71.87)}&  54.34 \scriptsize{(26.42)} & 219.76 \scriptsize{(39.26)} & 46.24 \scriptsize{(13.89)}  \\
		\vspace*{.1cm}
		{\small $n=500$}\\
		\vspace*{.1cm}
		S1 &  74.33 \scriptsize{(28.19)}& 6.77 \scriptsize{(4.06)} & 61.28 \scriptsize{(23.12)}& 5.21 \scriptsize{(3.17)} & 138.26 \scriptsize{(29.97)} & 23.19 \scriptsize{(7.55)}  \\
		\vspace*{.1cm}
		S2   & 85.29  \scriptsize{(24.09)}& 7.52 \scriptsize{(3.79)}& 105.95 \scriptsize{(65.39)}&  14.90 \scriptsize{(17.89)} & 132.39 \scriptsize{(24.02)} & 16.51 \scriptsize{(4.56)}   \\
		\vspace*{.1cm}
		S3   & 183.84  \scriptsize{(29.72)}& 44.09 \scriptsize{(8.45)}& 227.87 \scriptsize{(60.46)}&  60.98 \scriptsize{(23.32)} & 202.81 \scriptsize{(27.32)} & 58.94 \scriptsize{(6.25)}  \\
		\vspace*{.1cm}
		S4   & 136.39  \scriptsize{(20.17)}& 16.94 \scriptsize{(4.21)}& 161.04 \scriptsize{(75.70)}&  25.93 \scriptsize{(24.14)} & 143.91 \scriptsize{(20.08)} & 21.00 \scriptsize{(4.72)}  \\
		\bottomrule
	\end{tabular}
	\label{tab:1}
\end{table*} 

In this section we assess the performance of the proposed method and compare it with standard alternatives, namely  a location-scale DPM of Gaussians---the golden standard Bayesian nonparametric model for density estimation---and, for its close relations with our method, a SAPT. 

\begin{figure*}[t] 
	\centering
	\includegraphics[width=.79\textwidth]{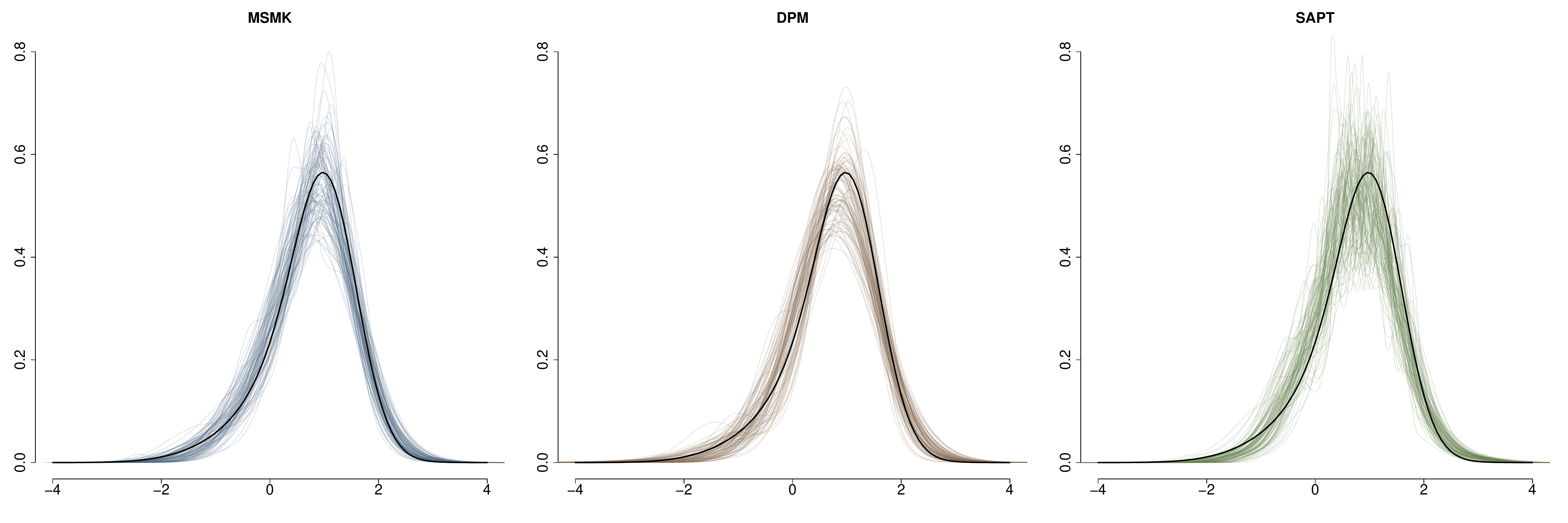}
	\includegraphics[width=.79\textwidth]{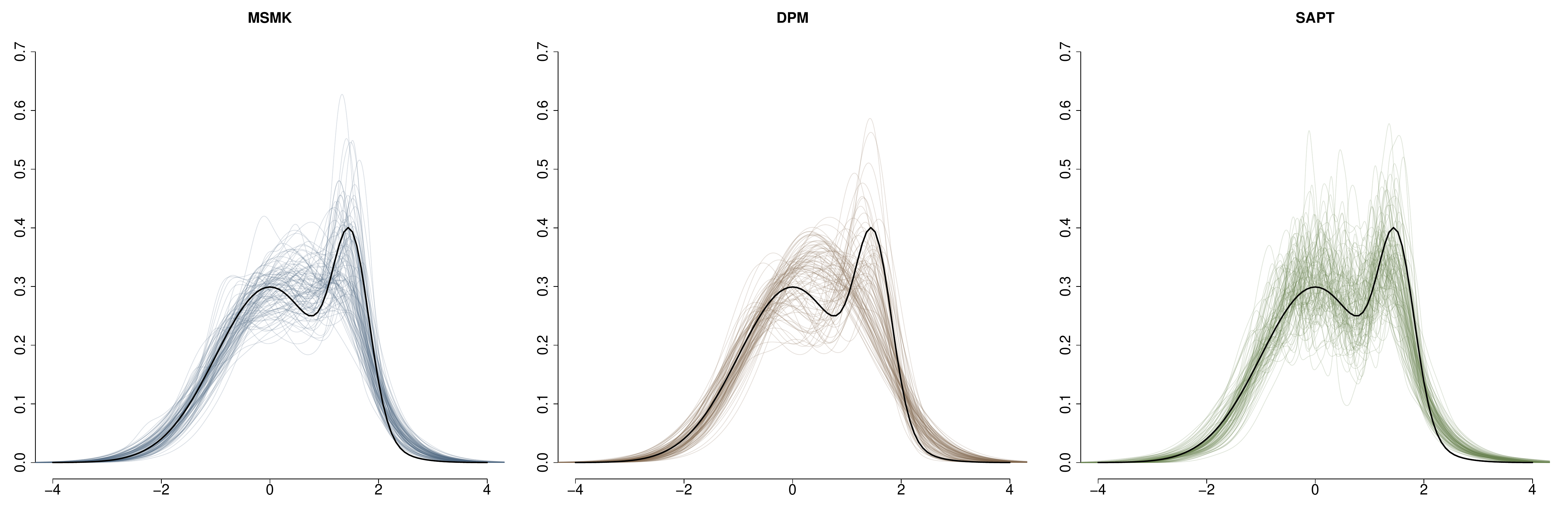}
	\includegraphics[width=.79\textwidth]{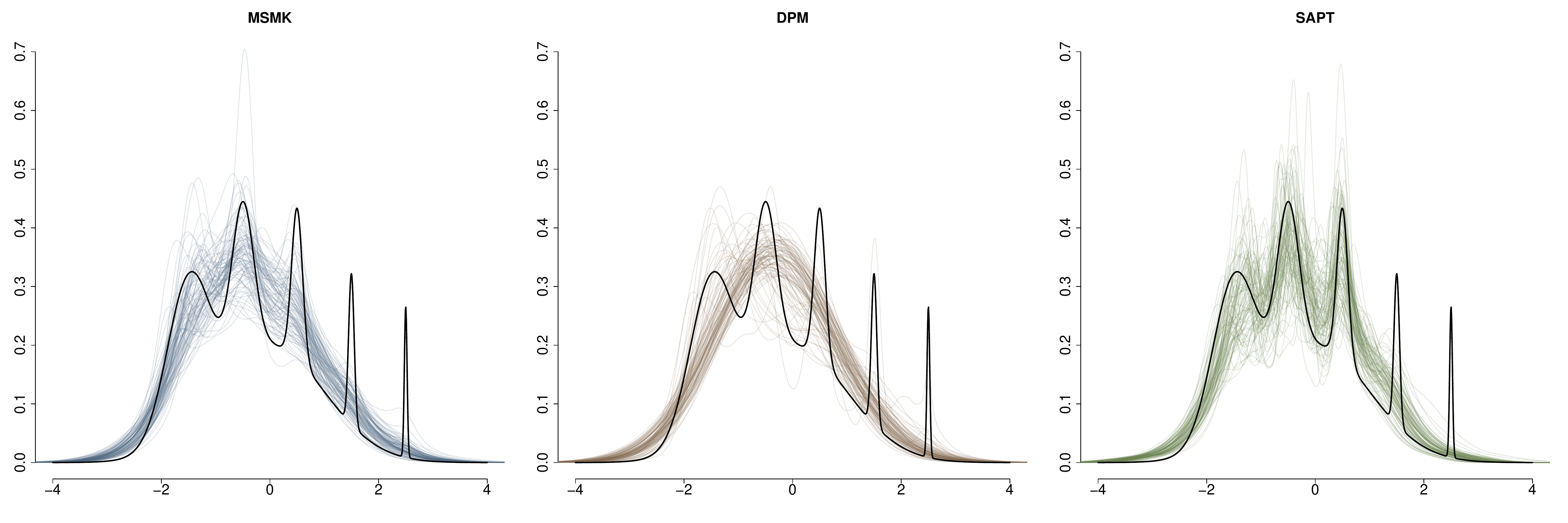}
	\includegraphics[width=.79\textwidth]{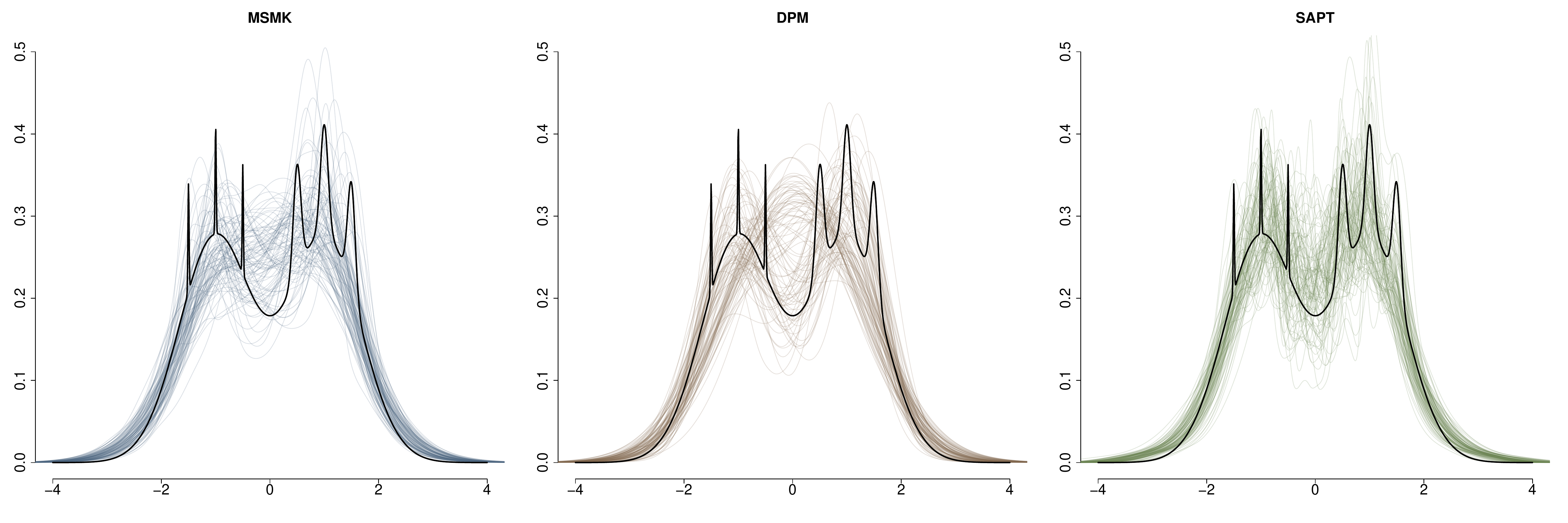} 
	\caption{Posterior mean densities (bright thin lines) for 100 independent samples (sample size $n=100$) and true densities generating the data (thick darker lines). Rows reports the results for the four scenarios. The results for MSM, DPM, and SAPT are reported in the first, second, and third columns, respectively. This figure appears in colour in the electronic version of the paper. }
	\label{fig:densim}
\end{figure*}

Synthetic data are simulated from different scenarios corresponding to varying degrees of global and local smoothness. As benchmark scenarios we used the densities reported in \citet{marron:1992}. For sake of brevity we report here the results for four scenarios (the results for all the densities of \citet{marron:1992} are reported in the Supplementary Materials) corresponding to a smooth unimodal skew density (S1), a smooth bimodal density (S2), and two densities with sharp local variability (S3 and S4). These densities are plotted with a thick dark line in Figure \ref{fig:densim}. Obviously, we expect the DPM to perform better in the first two scenarios---not having any multiscale structure---and our MSM and the SAPT to perform better in the last two scenarios. 
For each scenario we generate 100  samples of sizes $n=100$ and $n=500$.  Before fitting each model, data are standardized to have mean zero and variance one.  For the DPM we used the marginal P\'olya urn sampler implemented in the R package \texttt{DPpackage} \citep{jara:jss} while for SAPT for we used the Gibbs sampler described in the paper and implemented in set of R functions gently provided by the authors---that we thank warmly.  The performance of the three competing methods are evaluated in terms of $L_1$ distance and Kullback -Leibler (KL) divergence of the posterior mean densities from the true density evaluated on a grid of points.

For our MSM we set $G_0$ to be the standard normal distribution and $\lambda = k = 2^6$ for the inverse-gamma distribution $H_0$. This choice for $\lambda$ and $k$ leads to a high variance for the scale parameters reflecting mild prior information about these quantities. 
The maximum depth for the tree was set to $s'=6$. Consistently with the discussion on $\delta$ of the previous sections, we assumed  $\delta=0.5$. The value of $\alpha$ has been obtained numerically in order to match $\mathbb{E}(\tilde{S})=3$.  Finally, we set  $\beta=1.$
For the SAPT model we followed the specification presented in \citet{cipolli} and additionally let  $c \sim \text{Ga}(1, 1)$. The tree was grown up to $J = 6$ levels, consistently with the truncation induced in the multiscale stick-breaking. 
For the DPM the model specification is
\[
 f(\cdot) =  \int \phi(\cdot;\mu,\omega) d F(\cdot; \mu, \omega), \quad F \sim DP(\alpha,  F_0),
\]
with $F_0 =  \mbox{N}(m_1, \omega/\kappa) \times \mbox{IGa}(\nu_1,\psi_1)$ and additional hyperpriors 
\begin{align*}	
	& m_1  \sim \mbox{N}(m_2,s_2), \quad \kappa \sim \mbox{Ga}(\tau_1/2,\tau_2/2), \\
	&\psi_1 \sim \mbox{IGa}(\nu_2, \psi_2), \quad \alpha \sim \mbox{Ga}(a_0,b_0),
\end{align*}
with values of the parameters equal to $a_0=b_0=1,m_2=0, s_2=1,\nu_1=\nu_2=3, \psi_2=rs^2, r=0.1, \tau_1=2, \tau_2=200$ as in \citet{cipolli}.
Each MCMC algorithm was run for 1000 iterations  with a burn-in period of 200. 
Note that, contrarily to the prior specification for our MSM, for both the DPM and the SAPT we are adding different hyperprior distributions.

Table~\ref{tab:1} reports the results of the simulations study. Overall the performance of all the three methods are comparable. For $n=100$ our MSM of Gaussian performs sightly better---on average---both in terms of $L_1$ distance and KL divergence with respect to the competing methods. The lower values of $L_1$ distance and KL divergence attained by our MSM, are also coupled with less Montecarlo variability.
For the higher sample size of $n=500$, all the methods improve in terms of precision with our MSM always performing slightly better---but in the first scenario where, as expected, the DPM achieves the best performance. These results show that our MSM approach is able to adapt to the actual smoothness of the density and its performance make it a serious competitor of standard methods not only in  the situations where a multiscale structure is expected, but also 
when the density of the data is reasonably smooth.

Figure~\ref{fig:densim} gives additional insights on the results summarized in Table \ref{tab:1}. Each subplot of Figure~\ref{fig:densim} depicts, with thin bright lines, the posterior mean densities for each simulated  data sets (of size $n=100$) with different subplots in the same row denoting the three different competing methods. While the performance  in terms of $L_1$ distance and Kullback-Leibler divergence are most of the time comparable with the other methods, it is evident that for some specific data sets, the DPM estimates is a unimodal density, oversmoothing the true underlying density---see the second, third and fourth lines. 
On the other side, the SAPT  estimates avoid this oversmoothing but exhibit a very high variability with different data sets leading to estimates with prominent differences. The estimates obtained with our proposed method, instead, provide a better compromise between bias and variance, resulting in better posterior estimates that are smooth but also able to capture, abrupt local changes in the density---if present. Qualitatively similar results are also noticeable for the data sets with sample size $n=500$. See the Supplementary Materials for details.

\subsection{Roeder's galaxy speed data} 
\label{sec:galaxy1}

As benchmark data set to assess the performance of our method we use the famous Galaxy velocity data set of \citet{roeder1990} reporting the velocity of 82 galaxies sampled from 6  conic sections of the Corona Borealis.  Our goal here is to achieve comparable results in terms of goodness of fit with respect to standard methods that already showed to provide meaningful results---namely the DPM and allied models---as we do not expect a prominent multiscale structure.

\begin{figure}[h] 
	\centering
	\includegraphics[width=.4\textwidth]{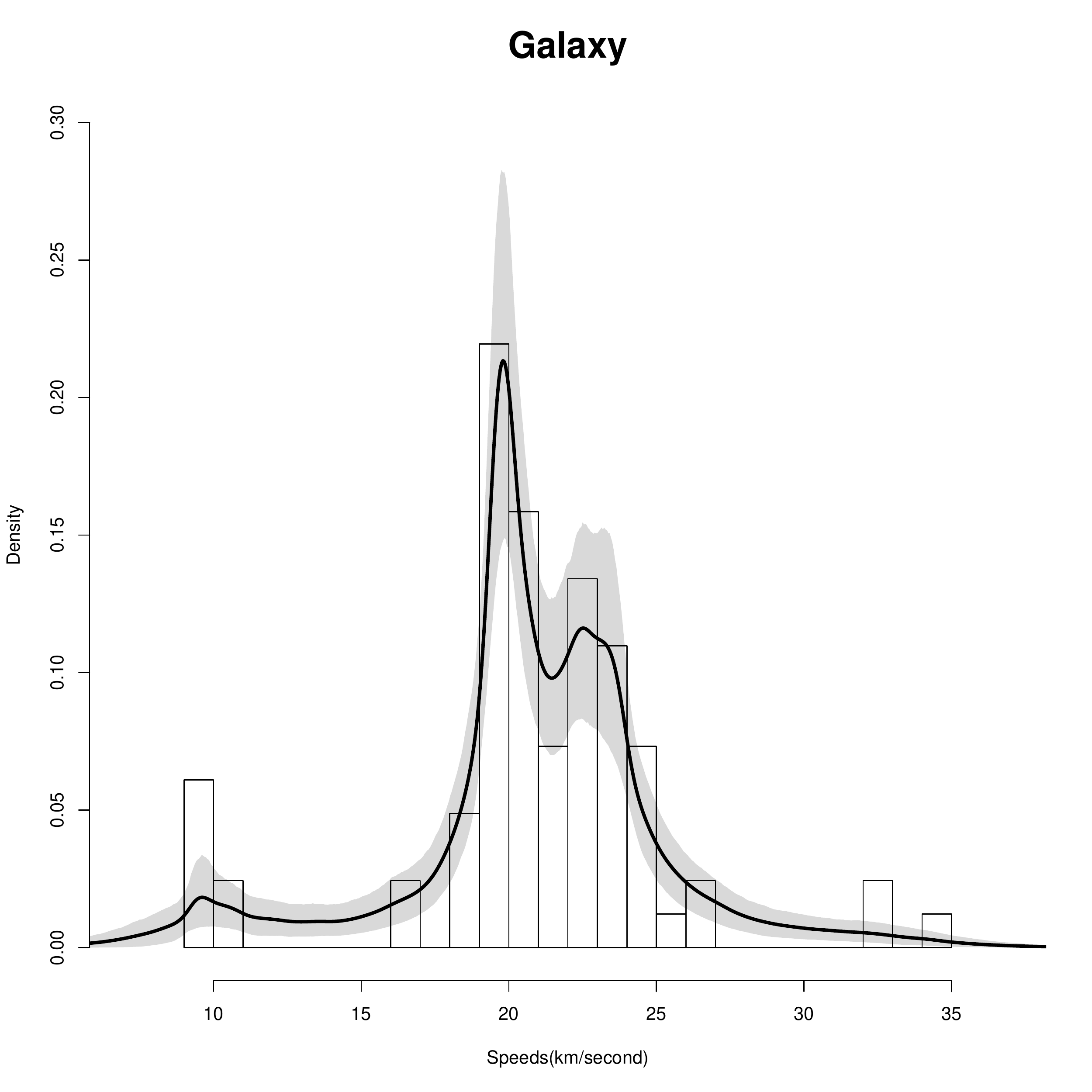} 
	\caption{Galaxy velocity data histogram and posterior mean density with 95\% posterior credible bands for the multiscale mixture of Gaussian model.
	}
	\label{fig:galaxy1}
\end{figure}

We used the same prior specification of the previous section but used a more conservative truncation of the binary trees, namely $s' = 8$. Figure  \ref{fig:galaxy1} depicts  the posterior mean density along with the histogram of the raw data and 95\% credible bands. As expected the method has a comparable performance with respect to state of the art competitors and achieve a log-pseudo marginal likelihood value \citep{gelfanddey} of $-217$, comparable to that of the DPM ($-212$) and SAPT ($-215$).

\subsection{Sloan Digital Sky Survey data} 
\label{sec:galaxy2}

We  consider a second astronomical data set consisting of $n=24\,312$ galaxies, drawn from the Sloan Digital Sky Survey first data release \citep[see][for details]{SDSS}. The galaxies are partitioned into $25$ different groups \citep{balogh2004bimodal}, by combining the separation in $5$ groups for different luminosity and in $5$ groups by different density---the latter being a physical characteristic of the galaxy that does not need to be confused with a probability density function.

Our goal here is twofold. From the astronomical point of view, considering this partition of the data as fixed, we want to estimate the probability density function of the difference of ultraviolet and red filters ($U-R$ color) for each group. Secondly, we use this example to show the flexibility of the proposed approach in dealing with complex situations proposing a modification of the mixture model discussed in Section \ref{sec:mixgaussian}. Specifically, for each group $g = 1, \dots, 25$, we assume the multiscale mixture
\begin{equation}
\label{eq:shark}
f_{g}(y)= \sum_{s=0}^\infty \sum_{h=1}^{2^s} \pi_{s,h}^{(g)} \phi(y;\mu_{s,h},\omega_{s,h}),
\end{equation}
where each set of weights $\pi_{s,h}^{(g)}$ is assumed to be generated independently according to the multiscale stick-breaking process introduced in Section \ref{sec:stickbreak} and each group-specific density $f_{g}$ shares a common set of kernel's parameters. The idea of a shared-kernel model accounts for  the existence of common latent information shared between groups and allows for borrowing of information in learning the values of the kernel's parameters. See \citet{lock2015shared} for a related approach.

Posterior sampling under the extension \eqref{eq:shark} can be performed following the details of Section \ref{sec:computation} and considering the update of each group specific set of weights independently by simulating 

\begin{align*}
S^{(g)}_{s,h} &\sim \mbox{Be}(1-\delta + n^{(g)}_{s,h}, \alpha + \delta (s+1)  + v^{(g)}_{s,h} - n^{(g)}_{s,h}) \\
R^{(g)}_{s,h} & \sim \mbox{Be}(\beta+r^{(g)}_{s,h}, \beta + v^{(g)}_{s,h}  - n_{s,h} - r^{(g)}_{s,h} ), 
\end{align*}
where
$v^{(g)}_{s,h}$, $n^{(g)}_{s,h}$, and $r^{(g)}_{s,h}$ are defined consistently to $v_{s,h}$, $n_{s,h}$, and $r_{s,h}$ of Section \ref{sec:computation} but considering only the subjects preassigned to group $g$.

\begin{figure*}[t] 
	\centering
	\includegraphics[width=.95\textwidth]{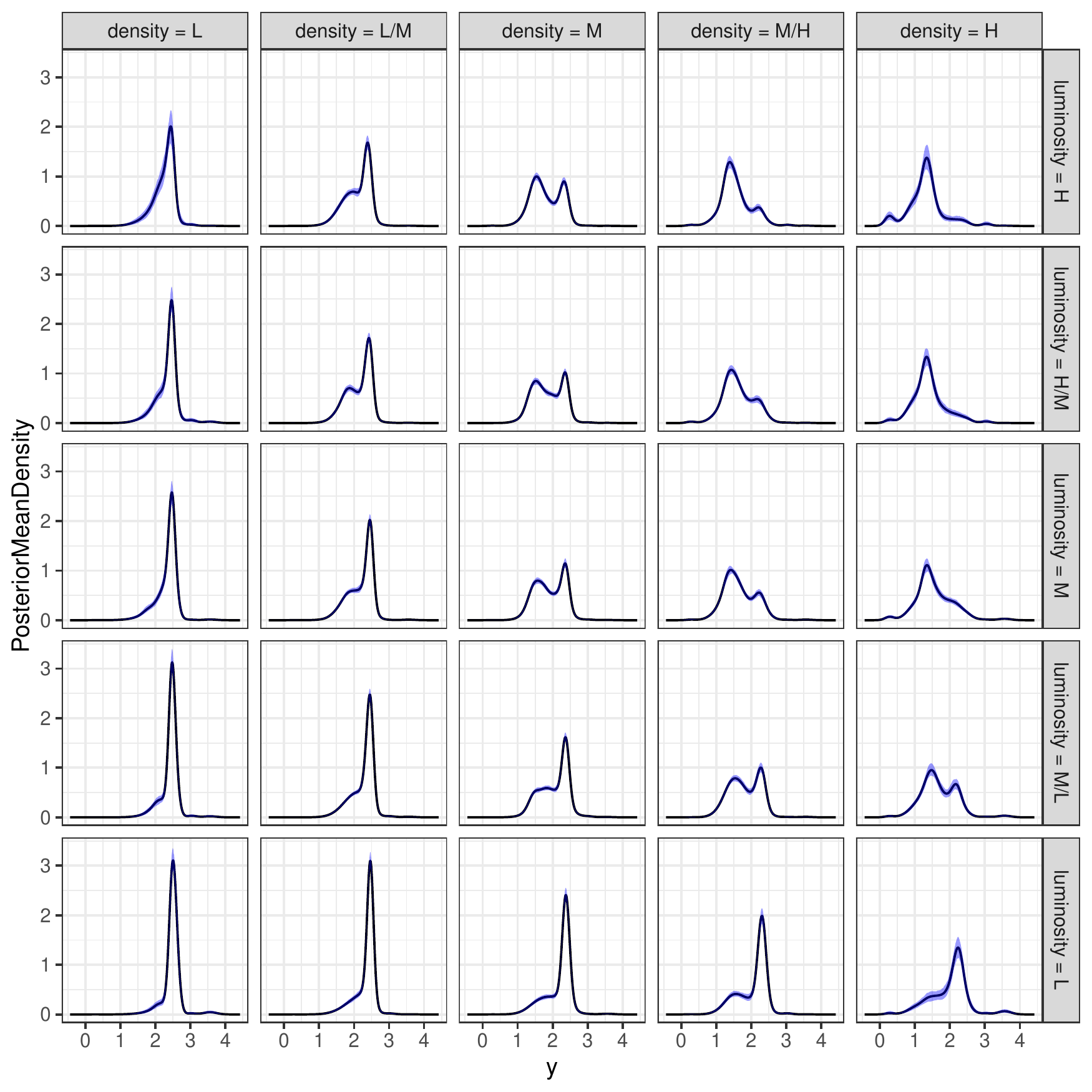}
	\caption{Sloan Digital Sky Survey data, $U-R$ color distributions, grouped with respect to luminosity and density. Black line: posterior mean densities; shaded areas: $0.95$ posterior credible  bands.}
\label{fig:sloan}
\end{figure*}

Assuming the same prior specification of the previous sections with $s'=4$ we run 1000 iterations of a Gibbs sampler with a burn-in period of 200. Figure \ref{fig:sloan} reports, for each group, the estimated posterior mean density along with 95\% credible bands. Many estimated densities show a clear bimodality which previous studies justified with the presence of two subpopulations of galaxies:  a blue and red population \citep{balogh2004bimodal}.  The different estimated densities clearly show different levels of global and local  variability that our model is able to capture. Notably, the posterior uncertainty is very low due to the joint effect of the moderately big sample size and of the shared kernel assumption.

\section{Discussion}
\label{sec:end}

	We introduced a family of multiscale stick-breaking mixture models for Bayesian n onparametric density estimation.  This class of models is made of two building blocks: a flexible multiscale stick-breaking process inspired by the PY literature and a stochastic process that generates a dictionary of stochastically ordered kernel densities. We showed that the $\delta$ parameter of the multiscale stick-breaking process---related to the discount parameter of the PY---makes the prior flexible and robust. Specifically it allows the method to achieve results comparable to those obtainable by more basic models endowed with an additional degree of hyperpriors--thus relieving the computational burden. The comparison with standard Bayesian nonparametric competitors showed, on average, superior performance in terms of finding the right smoothness of the unknown density. In addition, through the analysis of the Sloan Digital Sky Survey data, we also showed that the proposed formulation is amenable to extensions and generalizations to more complex settings involving hierarchical structures or covariates.

\section*{Acknowledgement}

The authors are supported by the University of Padova under the STARS Grant.

\section*{Appendix}
\begin{proof}[Lemma 1]
Let $\pi_s = \sum_{h=1}^{2^s} \pi_{s,h}$. For finite integer $S$, let $\Delta_S = 1 - \sum_{s=0}^S \pi_s$ which is equivalent to
\begin{align*}
\Delta_S & = \sum_{h=1}^{2^S} \Delta_{S,h}\\
& = 
\sum_{h=1}^{2^S} \left\{ (1 - S_{S,h} ) \prod_{r<S} (1-S_{r,\lceil h2^{r-s} \rceil}) T_{Shr},\right\}.
    \end{align*}
To establish the result, it is sufficient to show that the limit of each $\Delta_{Sh}$ for $S\to\infty$ is 0 a.s. Note that each $\Delta_{S,h}$ has the same distribution of
\[
\prod_{s=1}^S (1-S_s) T_{s-1},
\]
 with $S_s \sim \mbox{Be}(1-\delta, \alpha + \delta s)$ independent of $T_s \sim \mbox{Be}(\beta, \beta)$. 
Using Jensen's inequality
\begin{align*}
	\mathbb{E}[\log\{(1-S_s)T_{s-1}\}] \leq &	\log\{(1-\mathbb{E}[S_s])\mathbb{E}[T_{s-1}]\}\\ 
	= &\log\left(\frac{\alpha + \delta s + \delta}{2(\alpha + \delta s + 1)}\right) < 0,
\end{align*}
and therefore 
 \[
\sum_{s=0}^\infty \mathbb{E}[\log\{(1-S_s)T_{s-1}\}] =   -\infty.
\] 
Now use Lemma 2 of \citet{lancillotto} to obtain the result.
	\end{proof}

\begin{proof}[Lemma 2]
	\begin{eqnarray*}
&&\mathbb{E}[G(A)]  = \\
&& =  \mathbb{E}\left[   \sum_{s=0}^\infty \sum_{h=1}^{2^s} \pi_{s,h} \delta_{\mu_{s,h}} (A) \right] \\
&& =   
   \sum_{s=0}^\infty \sum_{h=1}^{2^s} \mathbb{E}\left[\pi_{s,h}\right]   G_0\left( A \cap  \Theta_{\mu;s,h} \right) 2^s \\	
&& =  
\sum_{s=0}^\infty \sum_{h=1}^{2^s} \frac{\left(1-\delta \right) \prod_{j=0}^{s-1} \left( \alpha +\delta (j +1)\right) }{\prod_{j=0}^s \left( \alpha +\delta j + 1 \right)}   G_0\left( A \cap  \Theta_{\mu;s,h} \right) \\	
&& =   
\sum_{s=0}^\infty \frac{\left(1-\delta \right) \prod_{j=0}^{s-1} \left( \alpha +\delta (j +1)\right) }{\prod_{j=0}^s \left( \alpha +\delta j + 1 \right)}  \sum_{h=1}^{2^s}  G_0\left( A \cap  \Theta_{\mu;s,h} \right) \\	
&& =   
G_0\left( A\right)  \sum_{s=0}^\infty \frac{\left(1-\delta \right) \prod_{j=0}^{s-1} \left( \alpha +\delta (j +1)\right) }{\prod_{j=0}^s \left( \alpha +\delta j + 1 \right)}    \\	
&& =   
G_0\left( A\right) 
	\end{eqnarray*}
\end{proof}

\bibliographystyle{apalike}
\bibliography{biblio}

\end{document}